\documentclass[aps,
prl,
english,
reprint]{revtex4-1}

\usepackage{amsfonts,amsmath,amssymb}
\usepackage{graphicx}
\usepackage[utf8]{inputenc}
\usepackage{hyperref}
\usepackage{babel}
\usepackage{enumitem}

\newcommand\e{{\rm e}}

\newcommand\be{\begin{equation}}
\newcommand\ee{\end{equation}}
\newcommand\bea{\begin{eqnarray}}
\newcommand\eea{\end{eqnarray}}

\begin{document}

\def\rhoo{\rho_{_0}\!} 
\def\rhooo{\rho_{_{0,0}}\!} 

\begin{flushright}
\phantom{
{\tt arXiv:2006.$\_\_\_\_$}
}
\end{flushright}

{\flushleft\vskip-1.4cm\vbox{\includegraphics[width=1.15in]{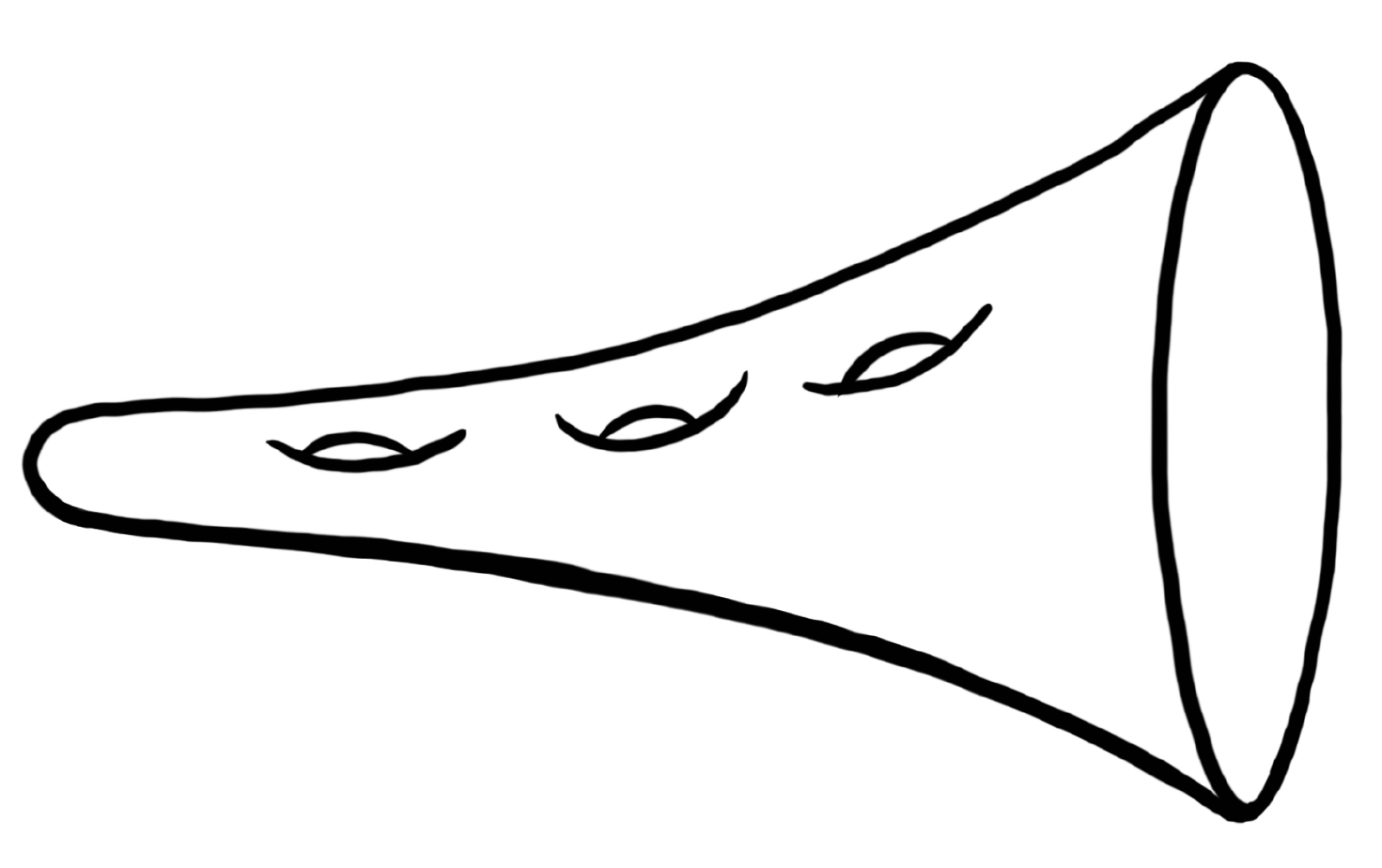}}}

\title{The Distribution of Ground State Energies in JT Gravity}
\author{Clifford V. Johnson}
\email{johnson1@usc.edu}

\affiliation{\medskip\\ Department of Physics, Jadwin Hall,  Princeton University,
Princeton, NJ 08544-0708, U.S.A.\vskip0.15cm}
 \affiliation{\medskip\\ Department of Physics and Astronomy, University of
Southern California,
 Los Angeles, CA 90089-0484, U.S.A.}


\begin{abstract}
It is shown that the distribution of the lowest energy eigenvalue of the quantum completions of Jackiw-Teitelboim gravity  is completely described by a non-linear ordinary differential equation (ODE)   arising from a    non-perturbative treatment of a special random Hermitian matrix model. 
Its solution matches the result recently obtained by computing a Fredholm determinant using quadrature methods. The new ODE approach allows for  analytical expressions for the asymptotic behaviour to be extracted. The results are highly analogous to the well-known Tracy-Widom distribution for the lowest eigenvalue of Gaussian random Hermitian matrices, which appears in a very diverse set of physical and mathematical contexts. Similarly, it is expected that the new distribution  characterizes a type of universality that can arise in various gravity settings, including black hole physics in various dimensions, and perhaps beyond. It has an  association to a special multicritical generalization of the Gross-Witten-Wadia phase transition. 

\end{abstract}

\keywords{wcwececwc ; wecwcecwc}

\maketitle


{\it Introduction.---} The Tracy-Widom distribution~\cite{Tracy:1992rf}  is ubiquitous, appearing in a wide variety of topics in theoretical and experimental physics, and mathematics. It  is, {\it e.g.}, the distribution of the edge eigenvalues of a Gaussian ensemble of large Hermitian matrices, the longest increasing subsequence of random permutations of integers~\cite{Baik:1998A},  height variances in certain  growth processes in the KPZ class~\cite{2004PhRvE..69a1103M}, and the onset of instability in certain random dynamical systems~\cite{May1972WillAL}.  It has become widely recognized as a new kind of universality. Notably, it can be described in terms of a special solution of the Painlev\'e~II ordinary differential equation (ODE), an equation that has far-reaching mathematical roots. The latter describes~\cite{Periwal:1990gf} the Gross-Witten-Wadia~\cite{Gross:1980he,*Wadia:2012fr,*Wadia:1980cp} third order phase transition in gauge theory, which helps strengthen the notion of universality for Tracy-Widom~\cite{Majumdar:2014}. The ODE description  is extremely useful since it allows for a characterization of the  asymptotic fall-off of  the distribution, useful for comparison to experimental data.

\begin{figure}[b]
\centering
\includegraphics[width=0.48\textwidth]{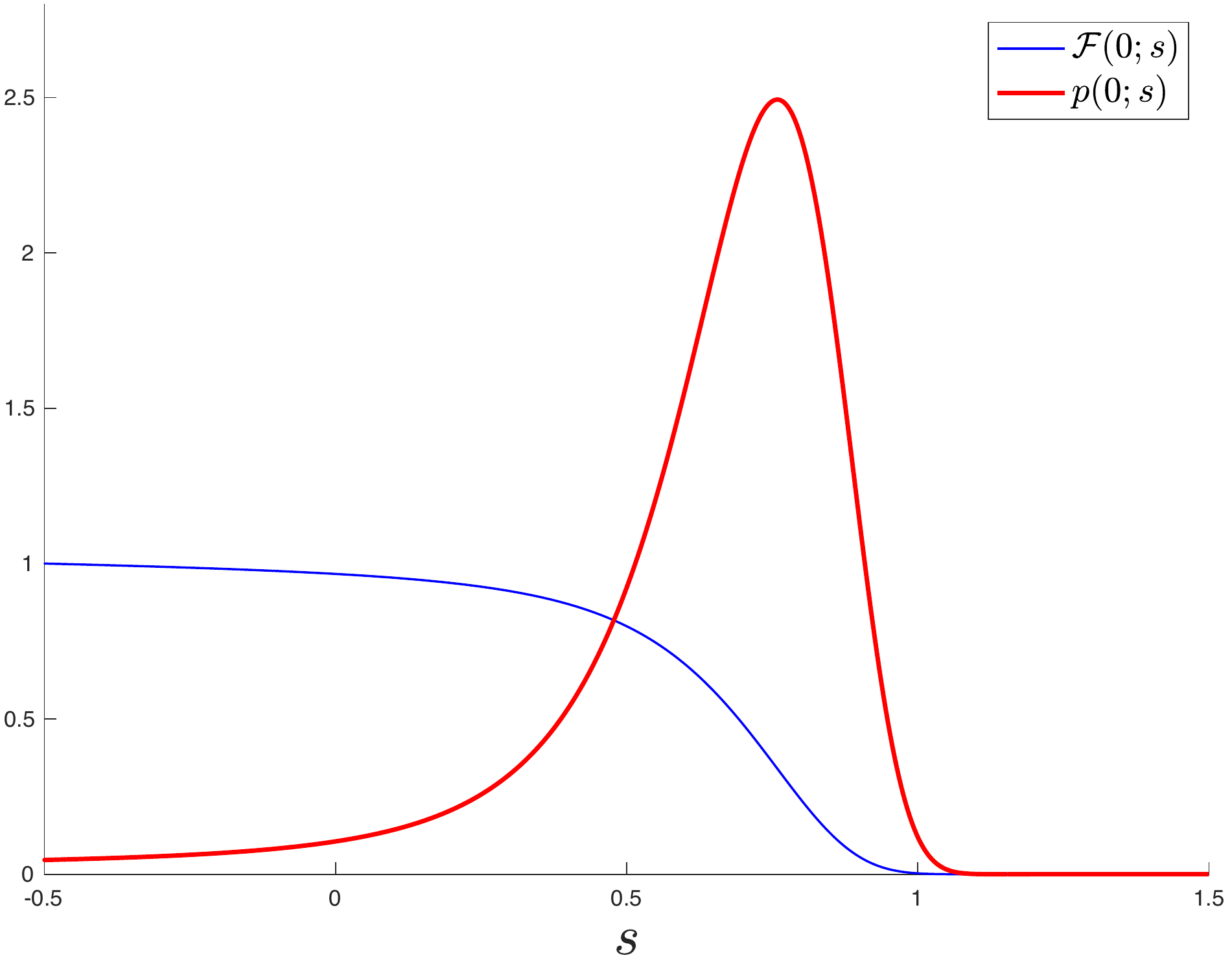}
\caption{\label{fig:JT-ground-state-distribution} The cumulative distribution function ${\cal F}(s)$, and probability distribution function $p(0;s)$ for the ground states (lowest energy) of the ensemble of JT gravity theories described by a random matrix model. Here,  $\hbar{=}{\rm e}^{-S_0}{=}1$.  }
\end{figure}

This Letter will derive and exhibit a new distribution, shown in figure~\ref{fig:JT-ground-state-distribution},  that arises in studies of  Jackiw-Teitelboim (JT)  gravity~\cite{Jackiw:1984je,*Teitelboim:1983ux}, a two dimensional  model that is also ubiquitous, arising in studies of the low temperature near-horizon dynamics of a variety of black holes in various dimensions. The distribution describes  the possible ground states of quantum completions of the model, and it will be shown  to emerge from a hierarchy of differential equations in a special limit. (They have been identified as a Painlev\'e XXXIV hierarchy, which also has a useful formulation as a kind of Painlev\'e II hierarchy.) 

\begin{figure}[b]
\centering
\includegraphics[width=0.48\textwidth]{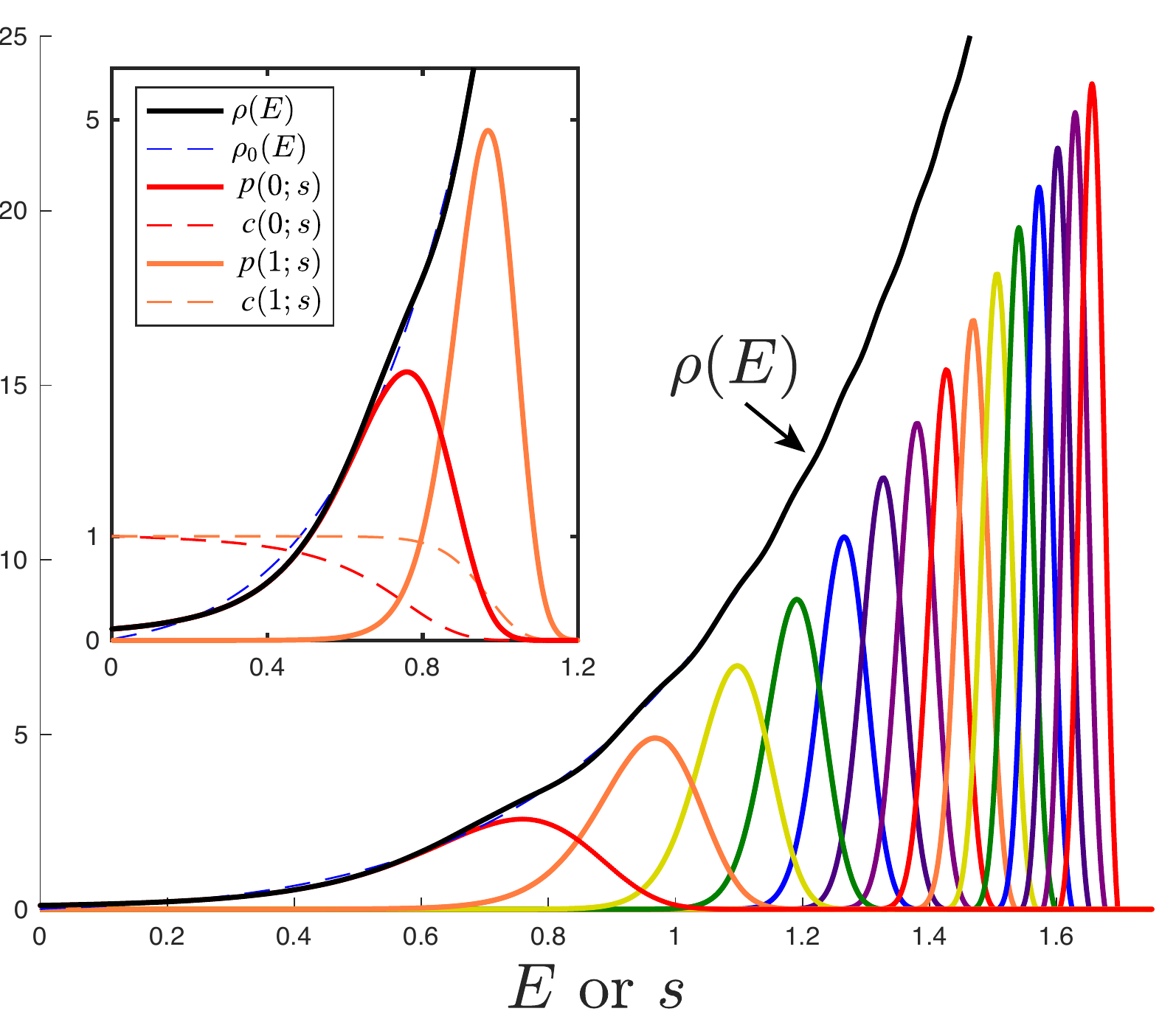}
\caption{\label{fig:JT-spectrum-from-fredholm} The full matrix model spectral density $\rho(E)$ (solid black), leading density $\rhoo(E)$ (blue dashed), and probability densities (also cumulative probabilities, dashed)  of the first 15  states of the  JT gravity microstate spectrum. ($s$ is another name for $E$ that will be explained later.)
Inset: Close-up of $\rho(E)$ and  the two first distributions. Here $\hbar{=}{\rm e}^{-S_0}{=}1$. }
\end{figure}

Quantum completions  of JT  gravity  have a discrete underlying microstate spectrum consistent with the finiteness and temperature dependence of the Bekenstein-Hawking~\cite{Bekenstein:1973ur,*Hawking:1976de} entropy. Such spectra are implicit in the    work of Saad, Shenker and Stanford~\cite{Saad:2019lba} that showed 
that the Euclidean quantum gravity path integral~\cite{Hartle:1976tp,*Gibbons:1976ue}  is perturbatively equivalent to the 't Hooft topological ($1/N$) expansion~\cite{'tHooft:1973jz} of a special random $N{\times}N$ Hermitian matrix model, but their nature becomes extremely clear in the fully non-perturbative treatments of refs.~\cite{Johnson:2019eik,Johnson:2020exp,Johnson:2021zuo,Johnson:2022wsr}, which make manifest the typical properties of the underlying  matrices. The point is that the gravity model's leading (at large~$E$) spectral density~\cite{Maldacena:2016hyu,*Jensen:2016pah,*Maldacena:2016upp,*Engelsoy:2016xyb}:
\be
\label{eq:schwarzian-leading}
\rhoo(E)=e^{S_0}\frac{\sinh(2\pi\sqrt{E})}{4\pi^2}\ ,
\ee
plus the infinite set of perturbative (in topology) corrections of ref.~\cite{Saad:2019lba} (an asymptotic series in $\hbar{=}\e^{-S_0}$, where~$S_0$ is the $T{=}0$ black hole entropy) is really an approximation to  an underlying {\it discrete} spectrum. The discreteness is invisible at any order of 
perturbation theory (large $E$), but is inevitable at lower $E$.
Formulating the model non-perturbatively (as an ensemble with a lower bound $\sigma$ on the  energy), refs.~\cite{Johnson:2021zuo,Johnson:2022wsr}  showed that the matrix model's full  spectral density is a discrete sum of probability peaks $p(n;E)$ for the allowed energy levels: 
\be
\label{eq:peaks}
\rho(E)=\sum_{n=0}^\infty p(n,E)\ .
\ee
Figure~\ref{fig:JT-spectrum-from-fredholm} shows the first 15 peaks for JT gravity (with $\sigma{=}0$), becoming increasingly narrow and densely packed on   returning to the large~$E$ (perturbative) regime.~\footnote{Ref.~\cite{Johnson:2021tnl} explains the incorporation of non-zero $\sigma$. Analogous peaks for several other types of JT gravity and JT supergravity were computed and presented in ref.~\cite{Johnson:2022wsr}.} 

It follows~\cite{Johnson:2022wsr} that the full matrix model {\it is not} JT gravity, but is better interpreted as the ensemble of possible quantum completions of the perturbative JT gravity. (This is  borne out by studying the thermodynamics of the full theory, and is consistent with holography.)
The matrix model's role should be understood precisely in the spirit of Wigner's~\cite{10.2307/1970079} original intent--not to give a definite spectrum, but providing a remarkable amount of data about the typical behaviour of the possible spectra. 

The distribution of possible ground states of the gravity system, $p(0;s)$, will be    this Letter's focus. 
The key result to report is that it is characterized by a non-linear ordinary differential equation (or more precisely, a family of them working in concert), whose properties will be exhibited and explored. The papers that first uncovered $p(0;s)$~\cite{Johnson:2021zuo,Johnson:2022wsr} used numerical quadrature methods (pioneered by Bornemann~\cite{Bornemann_2009}) to compute a certain Fredholm determinant  associated to the matrix model. Knowing its definition in terms of an ODE yields much more information about the distribution, however, such as analytic formulae for the asymptotic behaviour of its left and right tails. 
Moreover, the family of differential equations (a Painlev\'e~XXXIV hierarchy that maps to a Painlev\'e~II hierarchy) involved in the story will perhaps serve as a new bridge from black holes to other fields of physics and mathematics.

{\it The Differential Equation---}The derivation is straightforward, with all of the elements needed already present in the literature going back to the early 90s. 
The random matrix model is a study~\cite{Meh2004,*ForresterBook} of the ensemble of $N{\times}N$ Hermitian matrices $M$ with probability $P(M)=\exp(-N{\rm Tr}[ V(M)])/{\tilde Z}$ for polynomial $V(M)$,  fixed by matching to perturbative gravity. $V(M)$ can be chosen to be even. The physics can be described just in terms of the eigenvalues $\lambda_i$ of the matrices. In the large~$N$ limit they can be described with a smooth coordinate $\lambda{\in}\mathbb{R}$, and at leading order they form a distribution with  endpoints at some finite locations $\lambda{=}\pm\lambda_0$, described by some density function ${\tilde\rho}_0(\lambda)$ which is determined by~$V(\lambda)$. The universal physics arises from the behaviour  at an endpoint, say $-\lambda_0$. For the Gaussian case, ${\tilde\rho}_0(\lambda)$  describes the Wigner semi-circle law, vanishing as a square root at the end. The more general ``multicritical'' vanishing involving $k{-}1$ extra zeroes~\cite{Kazakov:1989bc,*Dalley:1991zs} will be of interest here too: ${\tilde \rho}(\lambda){\sim}(\lambda+\lambda_0)^{k-\frac12}$.    The focus is on the  neighbourhood of the endpoint, with  $E$ describing the scaled deviation of $\lambda$ away from $\lambda_0$ as $N{\to}\infty$. This is the ``double-scaling'' limit~\cite{Brezin:1990rb,*Douglas:1990ve,*Gross:1990vs,*Gross:1990aw}. There is an efficient description of the physics in terms of a family of orthogonal functions $\psi(E,x)$  (where $x\in\mathbb{R}$) determined by a Schr\"odinger problem ${\cal H}\psi(E,x){=}E\psi(E,x)$, with potential~$u(x)$, {\it i.e.}: ${\cal H}{=}{-}\hbar^2\partial_x^2+u(x)$. The parameter~$\hbar$ is the scaling piece of the $1/N$ expansion parameter that survives the limit, and  $u(x)$ is determined by $V(\lambda)$ through an ODE (the ``string equation"). (In the simple  Gaussian   ensemble, $u(x){=}{-}x$, and the $\psi(E,x)$ are Airy functions.)  A useful object is the   ``Kernel'':
\be
K(E,E^\prime)=\int_{-\infty}^0\psi(E,x)\psi(E^\prime,x)dx\ ,
\ee  in terms of which all questions about the ensemble can be answered. {\it e.g.,} the probability for the system to have  energy values  $\{E_i\}$, $i=1,2,\cdots m \leq N$ can simply be written as the determinant of the $m\times m$ matrix ${ K}(E_i,E_j)$. The  spectral density 
is the diagonal: $\rho(E){=}K(E,E)$. 

The core point is that knowing the function $u(x)$ is equivalent to knowing the matrix model potential, and everything can be determined henceforth. JT gravity can be captured {\it fully non-perturbatively} in this language~\cite{Johnson:2019eik}. 
The energy level probability peaks $p(n;E)$ of equation~(\ref{eq:peaks})  can be obtained by computing ``gap probabilities", also expressible  in terms of $K$. The probability of there being {\it no eigenvalues} on the energy interval $[a,b]$ is given~\cite{Gaudin1961SurLL} by the Fredholm determinant associated to the integral operator ${\bf K}_{[a,b]}$ that acts on functions of energy as follows: $\int_a^b f(E^\prime)K(E,E^\prime)dE^\prime{=}g(E)$. The probability is ${\cal F}(a,b){=}{\rm det}[\mathbf{I}-{\bf K}_{[a,b]}]$. This can be applied to the case of the lowest eigenvalue: The distribution of the ground states of the matrices $M$ in the ensemble comes from  {\it e.g.,}  $a{=}{-}\infty$ and $b{=}s$, a sample energy. Then ${\cal F}(s) $ is the cumulative distribution function (CDF) for the lowest energy, falling from unity at $s{=}{-}\infty$, certainty that there's no eigenvalue there, to zero at $s{=}{+}\infty$, since they all lie  on~$\mathbb{R}$. ($\mathbb{R}^+$ if there's a lower bound $a{=}\sigma$ on the ensemble). The transition between the two extremes is most rapid near the  distribution's edge. Writing ${\cal F}(s){=}\exp(-f(s))$, where $f(s)$ runs monotonically from zero to infinity as~$s$ sweeps over its range,  the probability distribution is: 
\be
\label{eq:peak-formula}
p(0,s)=-\frac{d {\cal F}}{ds} = \frac{df}{ds} \e^{-f(s)}\ .
\ee The latter form can be viewed as the probability of the energy  $E{=}s$ multiplied by the probability  that no energies have so far appeared at smaller $s$.
All that is needed is the function $f(s)$. 

For the {\it largest} eigenvalue of the Gaussian ensemble, (the energies can be anywhere on $\mathbb{R}$) Tracy and Widom showed that   $f^{\prime\prime}(z){=}q^2(z)$ where $q(z)$ satisfies  Painlev\'e~II: 
\be
\label{eq:painleve-II-A}
\hbar^2 q(z)^{\prime\prime}-zq(z)-2q^3=0\ ,
\ee  with 
$q(z){\to}\sqrt{-z}$ as $z{\to}{-}\infty$ and $q(z){\to}0{+} {\rm Ai}(z\hbar^{-\frac23})$ as $z{\to}{+}\infty$. (In their units $\hbar{=}1$, but it will be left more general here in preparation for the gravity case.)  This uniquely characterizes  $q(z)$ as  the  Hastings-McLeod solution~\cite{Hastings1980}. Exchanging $z{\to}\,{-}s$ adapts this to the case of the lowest eigenvalue. As preparation for  the gravity case, it is prudent to reformulate Tracy-Widom somewhat. This is readily done using equations that were derived for random matrix models and extensively studied at almost the same time as Tracy-Widom, amusingly~\cite{Morris:1990bw,Dalley:1992qg,Dalley:1992yi,Johnson:1992wr,Johnson:1992pu,Dalley:1992br}. What follows will be a very simple derivation of the overall structure, with some new elements. (Ref.~\cite{Nadal_2011} presents the basic idea, apparently  unaware that the core equations had already been derived long before. See a further note~\footnote{The equations presented give the entire generalization to any $k$. Note that the mathematical physics literature has  re-discovered these same  equations and relationships as part of studies of  B{\"a}cklund transformations of Painlev\'e~II equations~\cite{clarkson1999backlund}, and also generalizations of Tracy-Widom~\cite{claeys2010higher,claeys2012numerical,Akemann:2012bw}.} 
on the  literature.) Defining ${\cal R}[u]{=}u{+}x$, the equation:
\be
\label{eq:big-string-equation}
(u-s){\cal R}^2-\frac{\hbar^2}2{\cal R}{\cal R}^{\prime\prime}+\frac{\hbar^2}4({\cal R}^\prime)^2=0\ ,
\ee
can be interpreted as the string equation for a Gaussian ensemble of Hermitian matrices whose lowest eigenvalue is given by $E{=}s$.  
Next, notice that the associated Schr\"odinger equation at energy $s$, {\it i.e.,} $({\cal H}{-}s)\psi(s,x){=}0$,  can be written in a factorized form: 
\be
\label{eq:miura-map}
(\partial_x+v(x))(\partial_x-v(x))\psi(s,x)=0\ ;\quad 
v^2+\hbar v^\prime+s=u\ ,
\ee  which gives a very useful form for the wavefunction: $\psi(x){=}A\exp{(\frac1\hbar\int^x v(x^\prime) dx^\prime)}$, where $A$ is a normalization that will turn out to be equal to $1/\sqrt{2}$. So if  $v(x)$ is known, the wavefunction can be constructed, and hence  the probability $df/ds$, and so $f(s)$, can be computed. Well, it was noticed long ago~\cite{Dalley:1992br} (for the case $s=0$) that if $u(x)$ satisfies (\ref{eq:big-string-equation}) then the  transformation~(\ref{eq:miura-map}) from $u(x)$ to $v(x)$ (the ``Miura transformation''~\cite{doi:10.1063/1.1664700} in the  integrable model literature) gives  the  Painlev\'e~II equation (in a different form than before; an additive constant is present). It is equivalent to a rather succinct form for $v(x)$ in terms of $u(x)$ that will be useful shortly. For non-zero $s$ the result is, writing ${\cal S}{=}\frac{\hbar}{2}(u{-}s)^\prime{-}v(u{-}s)$:
\bea
\label{eq:painleve-II-B}
\label{eq:painleve-II-hierarchy}
&&v(x) =\frac{\hbar}{2}\frac{{\cal R}^\prime[u]}{{\cal R}[u]}\ ; \qquad  {\cal S}[v(x)]-x v(x)+\frac{\hbar}{2}= 0\ , \\
&&\hskip-0.5cm
\Longrightarrow\quad  \frac{\hbar^2}{2} v(x)^{\prime\prime}-(x+s)v(x)-v^3+\frac{\hbar}{2}=0\ . \nonumber
\eea
(Note the $s v(x)$ term, yielding a  more general form used recently in this context in ref.~\cite{Johnson:2020lns}.) This equation at $s{=}0$ was used in ref.\cite{Johnson:2019eik} to study the value of the spectral density at zero energy, but this can be done for general $E{=}s$. In fact a nice feature  is that, using  the first form in equations~(\ref{eq:painleve-II-B}),  the integral below~(\ref{eq:miura-map}) shows that 
the wavefunction is: 
$\psi(s,x){=}\frac{1}{\sqrt{2}}{\cal R}^\frac12$ ,
{\it i.e.}, the string equation itself is built out of the ground state wavefunction! So the density at $E{=}s$ is simply: 
\be
\label{eq:density}
\rho(s)=\int_{-\infty}^0\!\!\psi(s,x)^2 dx = \frac12\int_{-\infty}^0 {\cal R}[u(x,s)] dx=\frac{df}{ds}\ ,
\ee
where the last equality follows since the density {\it is} the probability of that energy, and the $x$ and $s$--dependence of ${\cal R}$ comes through its dependence on $u(x;s)$. That this is a re-writing of  the Tracy-Widom form follows from the fact that there is a {\it different} transformation~\cite{Morris:1990bw} that takes $u(x)$ to equation~(\ref{eq:painleve-II-A}): $u(x)=2^{1/3}(q^2(z)+z)$ where $ z=-2^{-1/3}(x-s)$, and then $df/ds = -\int_s^\infty q^2(z)dz$, and then $f^{\prime\prime}(s)=q^2(s)$, as before.

It is straightforward to state the generalization to the $k$th model. It  uses $ R_k[u]$, the ``Gel'fand-Dikii'' differential polynomials  in $u(x)$ and its derivatives, normalized so that the non-derivative part has unit coefficient: $R_k=u^k+\cdots+\#u^{(2k-2)}$ where $u^{(m)}$ means the $m$th $x$-derivative. {\it e.g.}, $R_1{=}u$, $R_2{=}u^2{-}\frac{\hbar^2}{3}u^{\prime\prime}$, $R_3 {=} u^3+{\hbar^2}(u^\prime)^2/2+{\hbar^2}uu^{\prime\prime}+{\hbar^4}u^{(4)}/{10}$. Successive $R_k$ can be obtained from a recursion relation~\cite{Gelfand:1975rn}. Now in the string equation~(\ref{eq:big-string-equation}),  putting ${\cal R} = R_k[u]+x$, gives~\cite{Dalley:1992qg,Dalley:1992yi,Johnson:1992wr} an infinite family of ODEs (later identified in the mathematical literature as a Painlev\'e~XXXIV heirarchy). Transformation~(\ref{eq:miura-map}) 
yields the $k$th generalization of the extended Painlev\'e~II equation on the RHS of equation~(\ref{eq:painleve-II-hierarchy}), now with ${\cal S}{=}S_k{=}\frac{\hbar}{2}R^\prime_k[u]{-}vR_k[u]$. Numerically solving it for $v(x)$ and integrating to get the wavefunction (for a given~$s$) can be done rather efficiently. 
This  readily produces curves for the $k$th generalization of Tracy-Widom.

This Letter's primary goal is to get new insights into the physics of JT gravity and black holes, and so  a further step is needed.  Refs.~\cite{Johnson:2019eik,Johnson:2020exp} showed explicitly   that a fully non-perturbative definition of JT gravity can be obtained from string equation~(\ref{eq:big-string-equation}) by studying a particular combination of {\it all} of the multicritical models, where ${\cal R}{=} \sum_{k=1}^\infty t_k R_k[u]+x$, with $t_k{=}\pi^{(2k-2)}/(2\,k!(k{-}1)!)$. At leading order in large $-x$ (where derivatives can be neglected), this pattern of $t_k$s results in the leading spectral density~(\ref{eq:schwarzian-leading}), and  corrections in $\hbar$ (or $1{/}|x|$) yield the matrix model perturbation theory of ref.~\cite{Saad:2019lba} that matches that of JT gravity.  A complete solution for $u(x)$ (with $u(x){\to}s$ for large $+x$)  defines the theory fully non-perturbatively. A key observation is that while having infinite numbers of $t_k$ contributing  gives  an infinite order differential equation, the contribution from successively higher $t_k$  gets rapidly smaller with $k$, and a truncation at high enough $k$ is enough to give very accurate results. Hence,  the construction offers a complete and explicitly computable results for questions about non-perturbative physics, some of which were already mentioned above. 

Now to complete  the story of the distribution. A truncation to $k{=}7$ works extremely well, yielding a 14th order highly non-linear ODE for $v(x)$. It  was relatively straightforward to solve it (as a boundary value problem) for $v(x)$, for several $s$ values, to good accuracy (see note~\footnote{They take  hardly more than 30s to 300s for larger, positive~$s$, and as little as just a few seconds for increasingly negative values, with reported error$\sim 10^{-5}-10^{-6}$.  This is one to two or more orders of magnitude smaller than the time taken for solving the $u(x)$ equation to the same accuracy.}). Each  $v(x)$  solution was used to construct $\psi(s,x)$, which then yielded 
$\rho(s)$ {\it via} equation~(\ref{eq:density}).  In this way, $f(s)$ could be readily extracted  to yield the smooth curve shown in figure~\ref{fig:JT-ground-state-distribution} (for the case of $\hbar{=}{\rm e}^{-S_0}{=}1$). The curve is normalized for the ensemble with lowest allowed energy $\sigma{=}{-}1$, by setting ${\cal F}(-1){=}1$.  Notice that this reproduces the form of the first peak in figure~\ref{fig:JT-spectrum-from-fredholm} obtained~\cite{Johnson:2021zuo,Johnson:2022wsr} by evaluation of a Fredholm determinant!

However, having the underlying ODE description allows for analytical expressions for   the peak's characteristic asymptotic fall-off behaviour.  This  is determined by studying the large $s$ behaviour of ${\cal R}$ in the negative~$x$ regime,  inherited from $u(x;s)$'s behaviour there. (See equation~(\ref{eq:density}).) There are both power law and exponential parts to the fall-off, but here the exponential parts will be the focus.
The  left asymptotic has the known ``instanton'' behavior of the matrix model spectral edge,  obtained by  {\it e.g.,} putting $u(x){=}u_0(x){+}\exp(-g(x))$ into equation~(\ref{eq:big-string-equation}) and assuming $g\gg g^\prime\gg g^{\prime\prime}$, {\it etc.}, and $u_0(x)$ is fixed by the leading solution at large negative $x$: ${\cal R}{=}0$.  It is $\sum_k t_k u_0^k+x{=}(2\pi)^{-1}\sqrt{u_0} I_1(2\pi\sqrt{u_0}){+}x=0$, where  $I_1(y)$ is the first modified Bessel function of $y$. Higher powers of $\hbar$   are dropped. The resulting $g(x)$ is of order $\hbar^{-1}$, and hence characterizes physics that is invisible at any order in small $\hbar$ perturbation theory. For example, for $k{=}1$, $u_0{=}\sqrt{-x}$ and the result is $g(x){=}2\sqrt{{-}(x{+}s)}/\hbar$. This gives, after integrating over the negative $x$ region, a dependence involving exponentials of $-\frac43[{-}(x{+}s)]^\frac32/\hbar$, yielding the  left asymptotic  fall-off  $~\exp(-\frac43(-s)^{\frac32}/\hbar)$ known for Tracy-Widom.   Generally, at small $\hbar$  the leading fall-off is $\exp({-V_{\rm eff}(s)})$ where the  effective potential for a single eigenvalue is $V_{\rm eff}(s){=}2\pi\int^{-s}\!\!\rho_0(E)dE$~\cite{David:1990ge}. For $k{=}1$,  $\rhoo(s){=}(-s)^\frac12/\hbar\pi$. For JT gravity, using~(\ref{eq:schwarzian-leading}), $V_{\rm eff}(s){=}(4\pi^3\hbar)^{-1}[\sin(2\pi\sqrt{-s}){-}2\pi\sqrt{-s}\cos(2\pi\sqrt{-s})]$~\footnote{This was done in  ref.~\cite{Saad:2019lba}. For a review of its derivation using WKB approach to the wavefunctions see ref.~\cite{Johnson:2021tnl}.}. This works down to  where $V_{\rm eff}$ first vanishes, signalling an instability. For larger $\hbar$, strong coupling effects modify $V_{\rm eff}(s)$,   extending the range further  left~\footnote{Some further illustration and discussion of this is  in the Supplemental Material, parts II and III.}. 

The leading right fall-off will give new  behaviour, valid for {\it any}~$\hbar$, notably. Seeing first  how it emerges  for Tracy-Widom is instructive. Notice  that for some  {\it positive} $s$ the function $g(x)$ vanishes in the negative $x$ regime (at $x{=}{-}s$ for this $k{=}1$ case) {\it i.e.}, the instanton approximation  above has become  invalid: The presence of positive $s$ becomes highly significant in the negative~$x$  regime beyond that point. So $u(x;s)$ (and hence ${\cal R}[u]$)  grows with~$x$  beyond that point, and ultimately since~$u{\to}s$ at large positive $x$, ${\cal R}[u]$ grows linearly with~$x$ as ${\cal R} {=} s+x$. For increasingly large $s$ this is the dominant behaviour for much of the region after $x{=}{-}s$, and so the integral~(\ref{eq:density}) yields $df/ds{=}s^2/4$. Hence, $f(s){=}s^3/12$, and the right exponential fall-off of the distribution is characterised by $\exp(-s^3/12)$, the precise result for the Tracy-Widom case.  It is faster (in $s$) than for the left edge, resulting from the fact that (in the Dyson gas picture)  an energy level moving deeper into the bulk of levels is pushed back more strongly by all the others. The same is true for the new distribution, as is evident in figure~\ref{fig:JT-ground-state-distribution}. The observations made for the $k{=}1$ case persist~\footnote{Some examples are shown in the Supplemental Material as illustration of this, along with some discussion pertaining to non-perturbative features seen for the left edge.}, with a rise to linearity for  ${\cal R}[u(x,s)]$ that is now ${\cal R}{=}\sum_k t_k s^k{+}x{=}C(s)+x$ where $C(s){=}(2\pi)^{-1}\sqrt{s} I_1(2\pi\sqrt{s})$.  For large $s$ this results in $df/ds{=} C(s)^2/4$, and hence an exponential fall off with argument given by  $f(s){=}\frac14\int C(s)^2 ds$,
which has a closed form in terms of modified Bessel functions, but not shown here.  
As already stated, there are also power-law parts of the fall-off. It would of course be of value  to do further analysis of the equations to characterize those.

{\it Closing Remarks.---}In the  Tracy-Widom case, the two asymptotic regions map to the strong and weak coupling  phases connected by  the Gross-Witten-Wadia~\cite{Gross:1980he,*Wadia:2012fr,*Wadia:1980cp} third order transition~\cite{Majumdar:2014}. The argument of the exponential fall-off in each regime maps to the leading part of the free energy in the different phases. This relation to a  phase transition helps motivate why it is so  universal.  For the new distribution presented here for quantum JT gravity ground states, there is again a map to  two different phases, now  connected by a very special transition built by combining higher order analogues of the GWW transition. This transition is worth further study. 



{\it Acknowledgments.---}CVJ  thanks Felipe Rosso for comments, the  US Department of Energy for support (\protect{DE-SC} 0011687), and  Amelia for her support and patience.    

\bibliographystyle{apsrev4-1}
\bibliography{Fredholm_super_JT_gravity1,Fredholm_super_JT_gravity2}

\vfill\eject

\appendix

\setcounter{figure}{0}
\renewcommand{\thefigure}{S\arabic{figure}}

\begin{widetext}
\section{\Large Supplemental Material}
Some additional figures and discussion are provided here as further illumination of some of the Letter's material.

\section{I. The $k{=}1$ (Tracy-Widom) case.}
Figure~\ref{fig:Supplement-A} (left)  shows some samples of the solution for $v(x)$, for $s=-2$ to $s=5$ in half integer increments. They were then used to construct the quantity $\psi(x)^2{=}{\cal R}(x)/2$, and figure~\ref{fig:Supplement-A} (right) shows the results. Crucial here for the large positive $s$ asymptotic of the distribution (as discussed in the last part of the Letter) is the fact that a significant rise begins in the negative $x$ region when $s$ is positive.  The rapid onset of linearity is evident from the curves, and in this approach it is responsible for the famous $\exp(-s^3/12)$ leading fall-off into the bulk for Tracy-Widom.

\begin{figure}[h]
\centering
\includegraphics[width=0.45\textwidth]{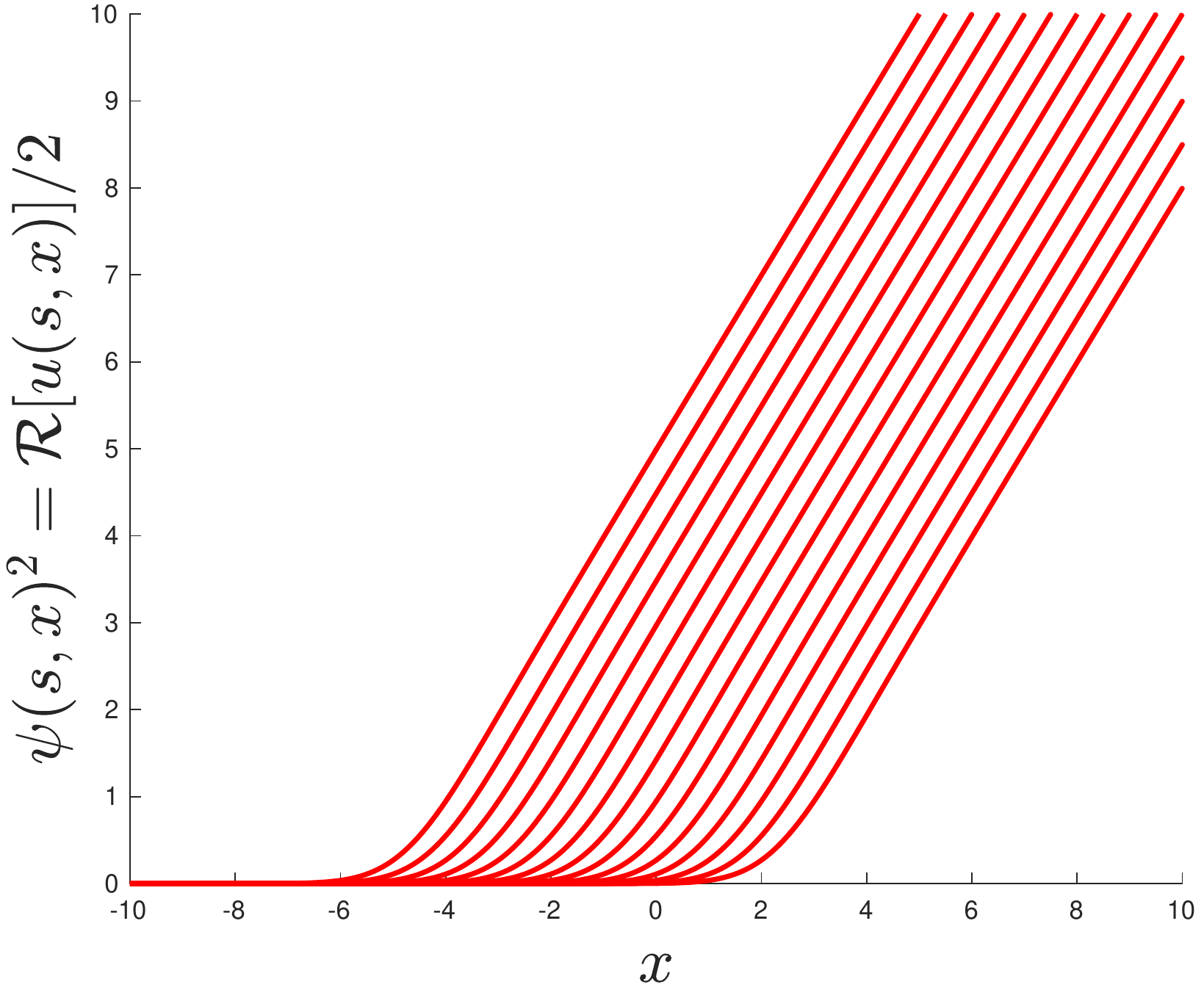}\hskip1cm \includegraphics[width=0.48\textwidth]{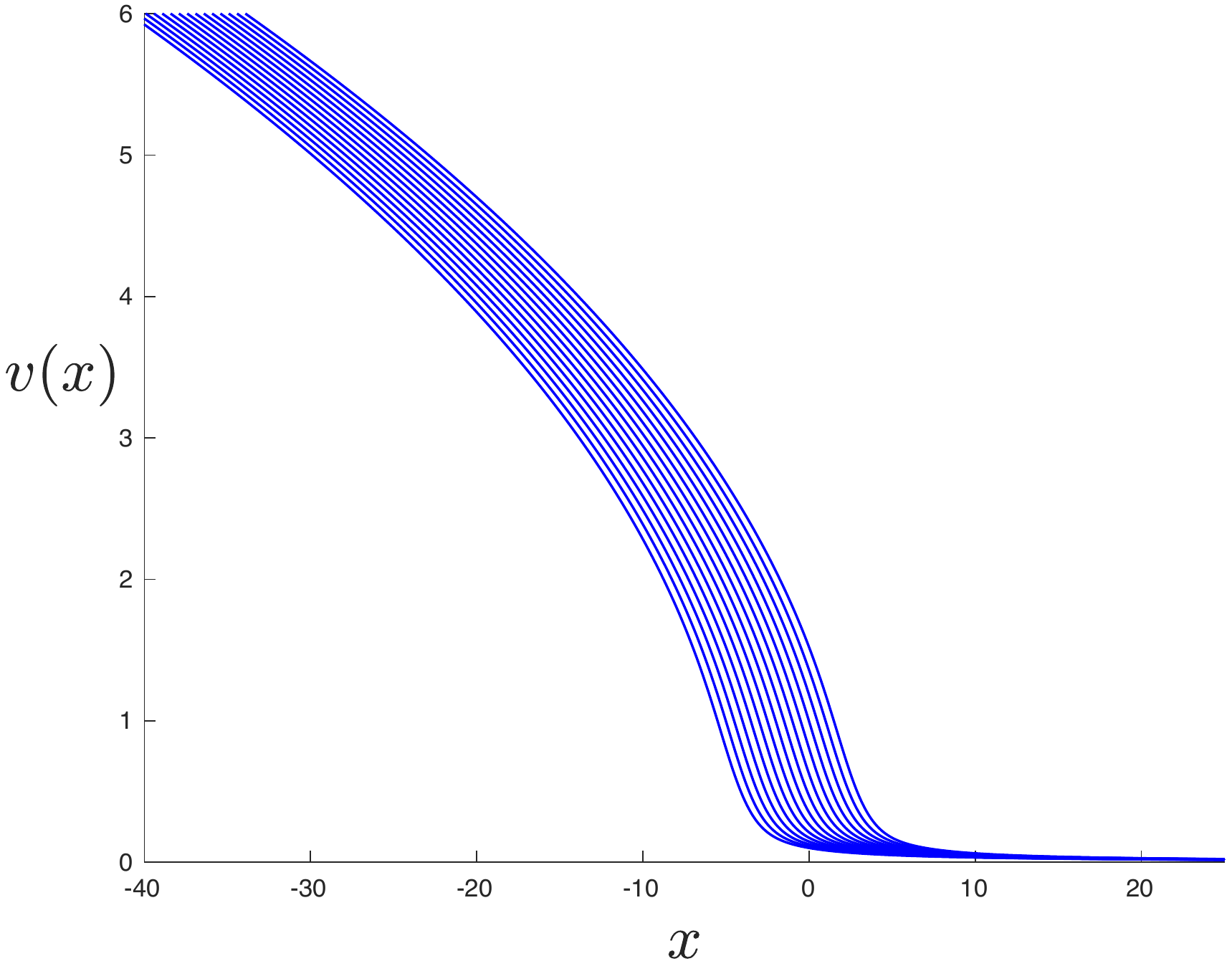}
\caption{\label{fig:Supplement-A}   }
\end{figure}

The Tracy-Widom curve that results from the previous results  is shown in figure~\ref{fig:Supplement-B} (left) for $\hbar=1$, and its analogue for $\hbar=\frac{1}{5}$ is shown on in  figure~\ref{fig:Supplement-B} (right) for comparison. Crucially, the left fall-off is swifter for the latter case, resulting in a narrower peak, and much smaller value of the probability/density at the origin. This is what should of course result from taking a step closer to the classical ($\hbar\to0$) limit where all peaks shrink to zero width and form a continuum. (The appropriate rescaling of the energy axis by $\hbar^{\frac23}$ further facilitates this comparison.)
\begin{figure}[h]
\centering
\includegraphics[width=0.45\textwidth]{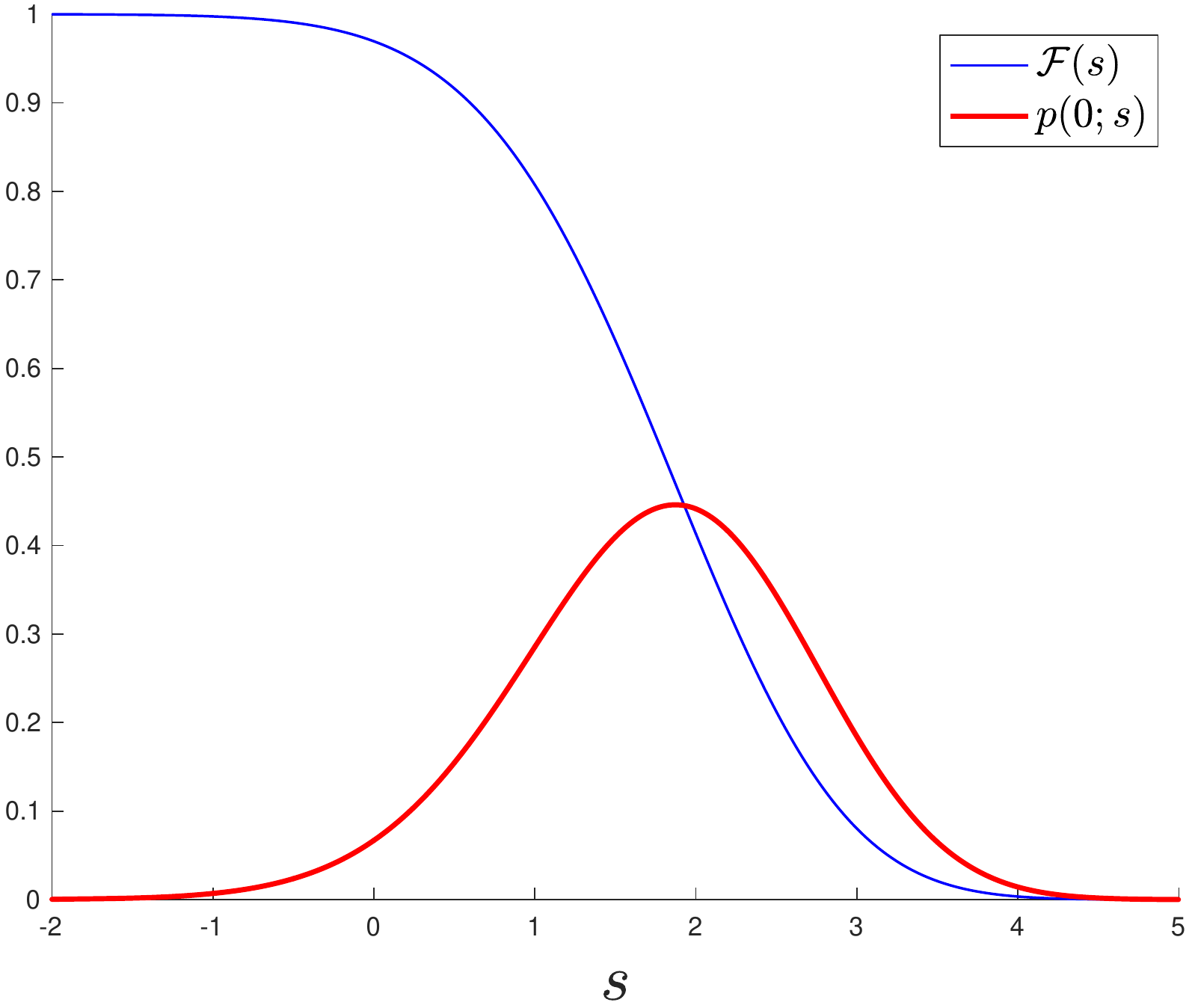}\hskip0.5cm \includegraphics[width=0.48\textwidth]{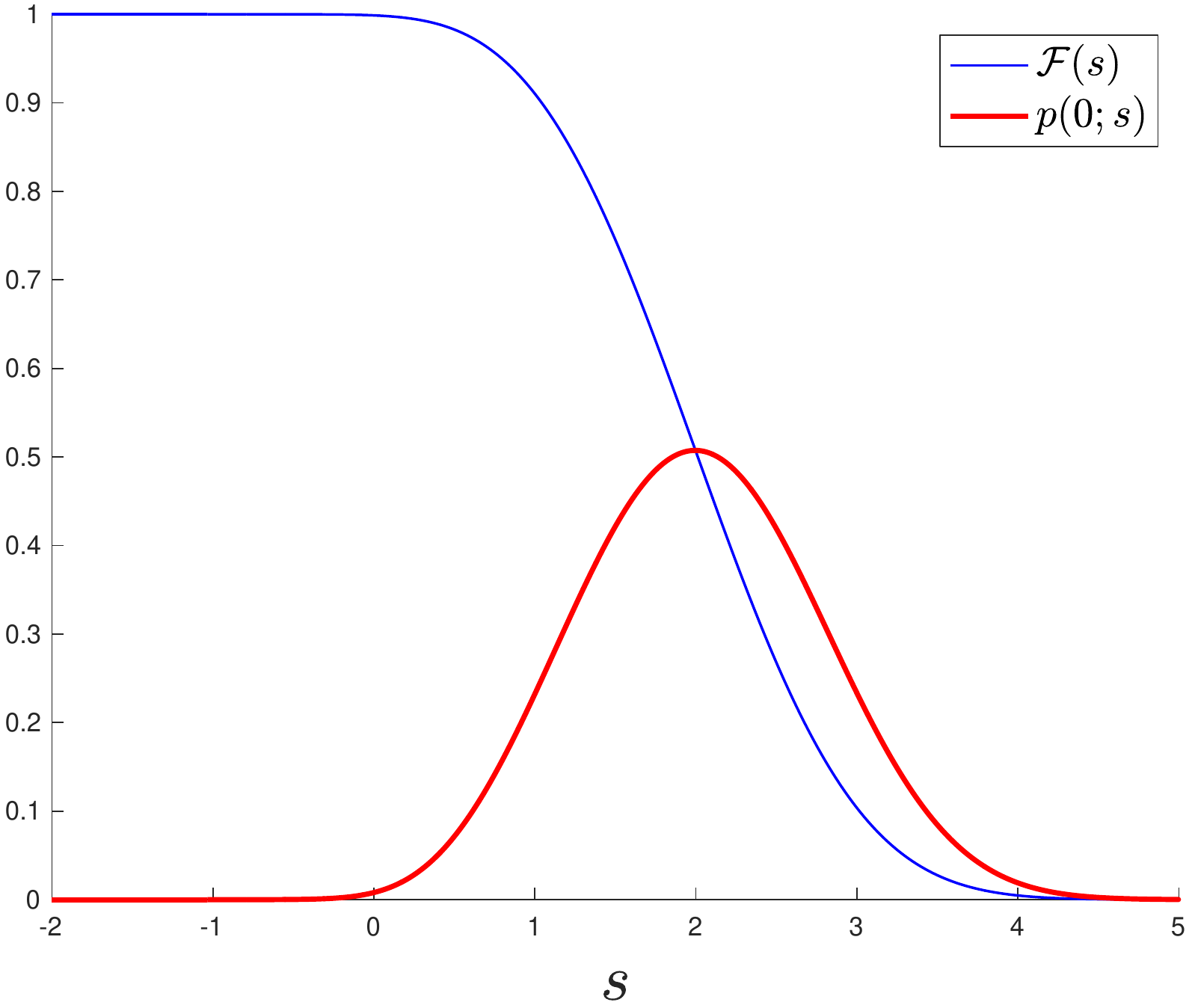}
\caption{\label{fig:Supplement-B} }
\end{figure}

\section{II. The $k{=}2$ case.}

As explained in the Letter, probability distribution curves are readily generated for any $k$, but this case (and any even $k$) has special features. A well-known statement about such cases, based on a semi-classical  analysis~\cite{David:1990ge}, is that they are ``unstable'' because eigenvalues can tunnel out of the bulk configuration because $V_{\rm eff}(s)\sim -s^{5/2}/\hbar$ is negative. Something should go wrong as soon as $s$ goes negative. The techniques developed here show precisely what that means, how the full non-perturbative results modify this expectation, and how to get meaningful results. (The core explorations of this were already done 30 years ago in ref.~\cite{Johnson:1992wr}, but the current discussion in terms of the distribution makes it more pleasingly explicit.) Figure~\ref{fig:Supplement-C} (left)  shows the distribution (for $\hbar{=}1$) down to $s=-0.4$, along with some sample curves for $v(x)$ in figure~\ref{fig:Supplement-C} (right). This is already in contradiction to the semi-classical expectation. What is going on is that for finite $\hbar$ there are  smooth solutions of the equation that have no continuous $\hbar{\to}0$ limit (see also ref.~\cite{Johnson:2021tnl}), whose role is to  extend the space of sensible solutions that are accessible with just the classical intuition. Differently put, strong coupling effects modify the semi-classical expectations, allowing for some reach into the negative $s$ regime. Studying the ensemble of random Hermitian matrices with lowest energy $\sigma=-0.4$ (or of course higher) will give  meaningful non-perturbative physics that connects to the perturbative results. 

\begin{figure}[h]
\centering
\includegraphics[width=0.45\textwidth]{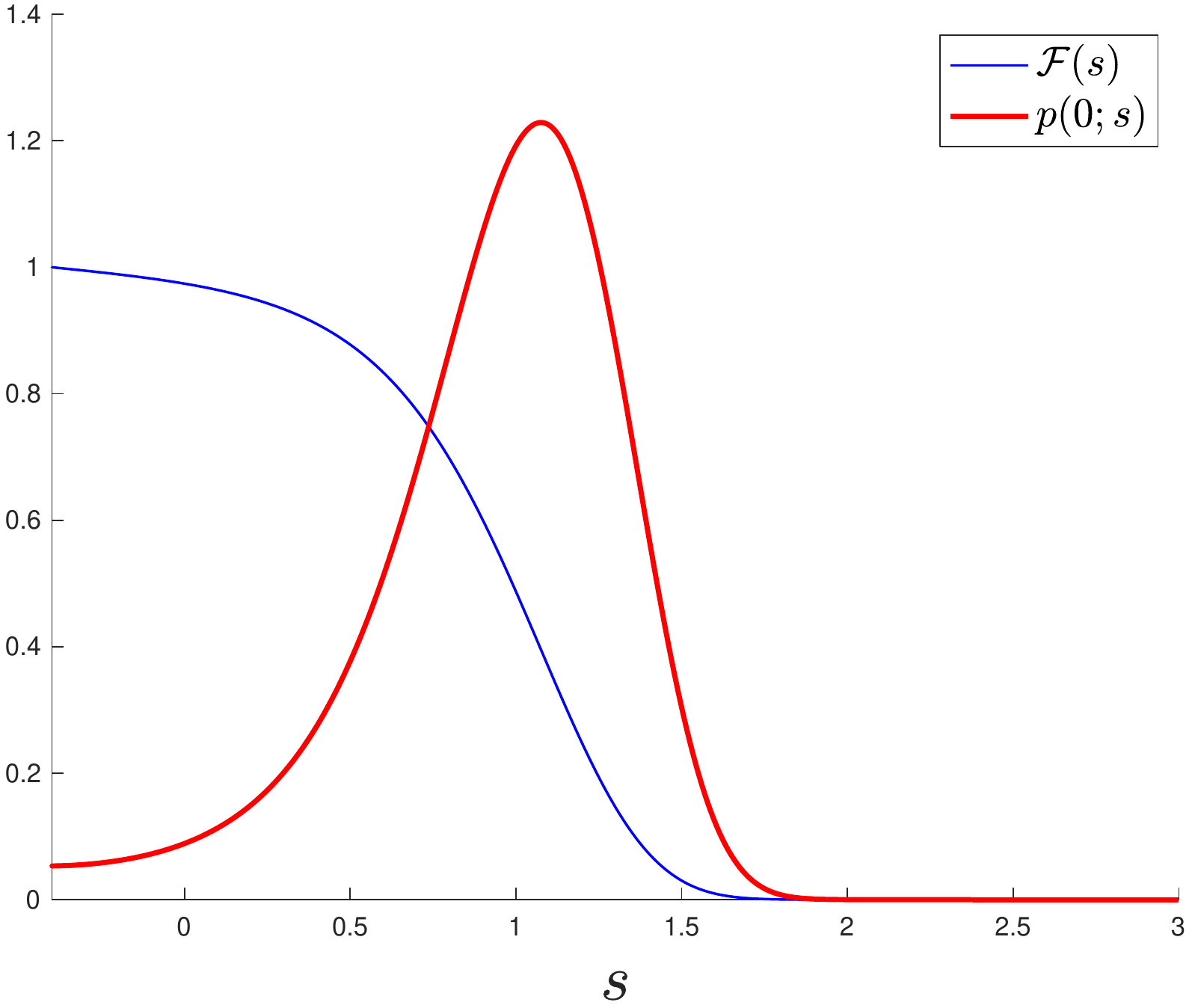}\hskip0.5cm \includegraphics[width=0.5\textwidth]{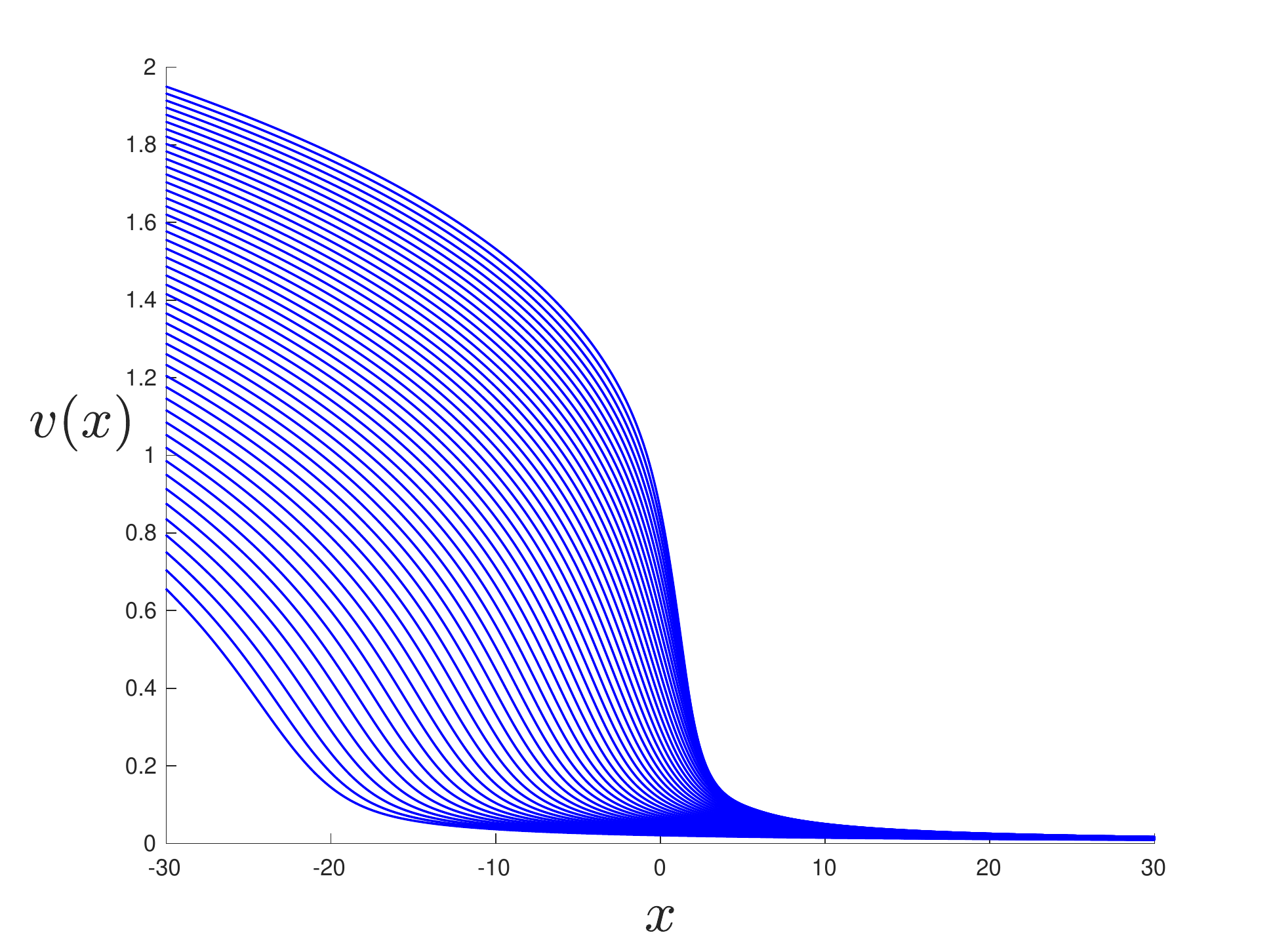}
\caption{\label{fig:Supplement-C}}
\end{figure}

However, proceeding to values much lower than this begins to show that the tail begins to rise again. See figure~\ref{fig:Supplement-D} (left). This is the result of the full strongly coupled  $V_{\rm eff}(s)$ going negative. Going to lower $s$ makes this rise more marked, at the expense of the original peak losing its height (conserving probability). In the limit of very large $-s$, the new peak grows increasingly high, representing the expected ``tunnelling-to-oblivion'' instability of the $k=2$ case when energies are allowed to be anywhere to the left. (This also fits with the fact that the string equation for $u(x)$  in this limit, Painlev\'e~I, has no smooth real non-perturbative solution for all $x$.) Figure~\ref{fig:Supplement-D}   (right) shows some of the $v(x)$ solutions, where some (controllable but marked) undulations develop at increasingly small $s$. These were observed for $u(x)$ long ago in refs.~\cite{Johnson:1992wr}, since of course $u(x)$ and $v(x)$ are related by the Miura map of the body of the Letter.

\begin{figure}[h]
\centering
\includegraphics[width=0.45\textwidth]{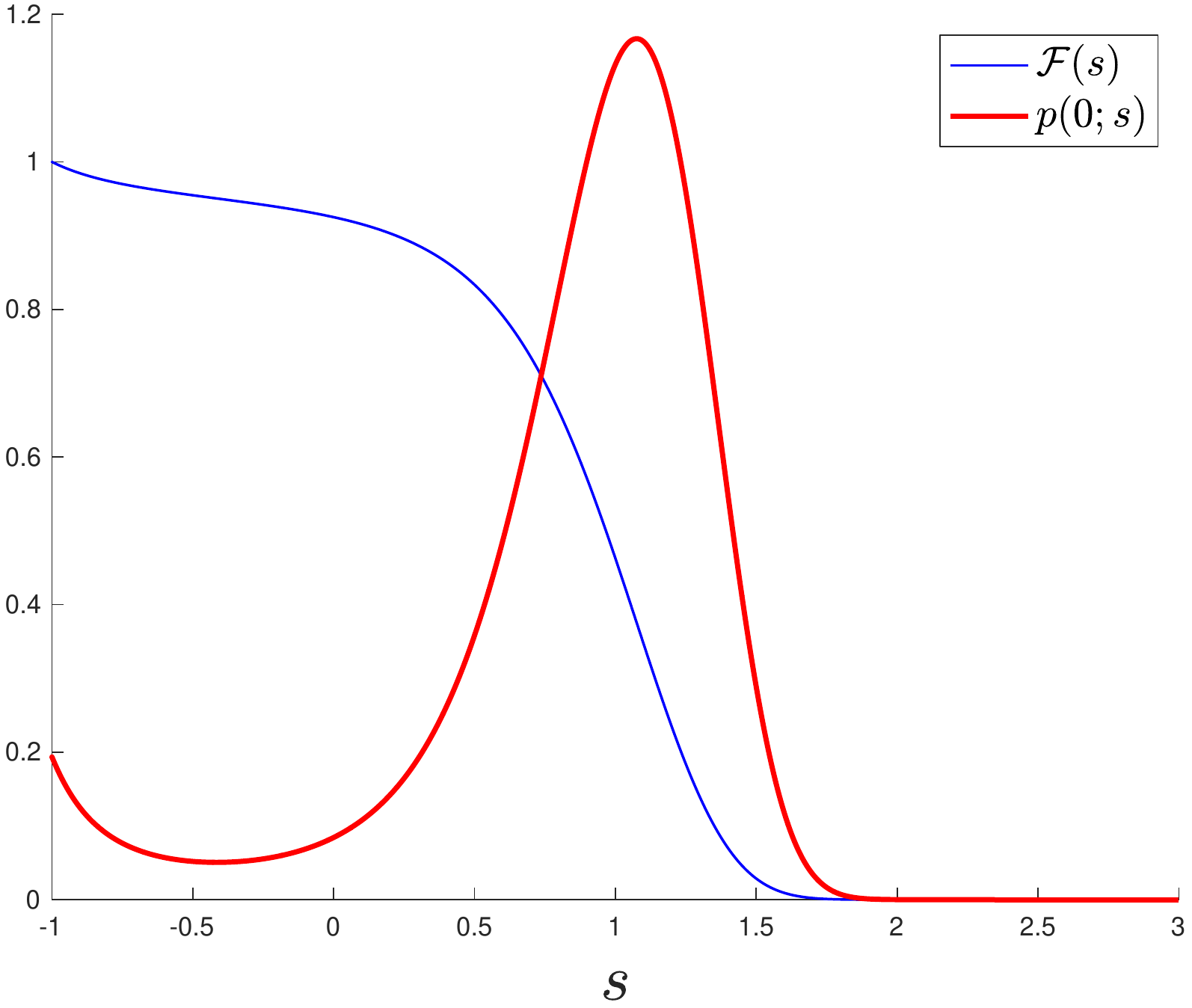}\hskip0.5cm \includegraphics[width=0.5\textwidth]{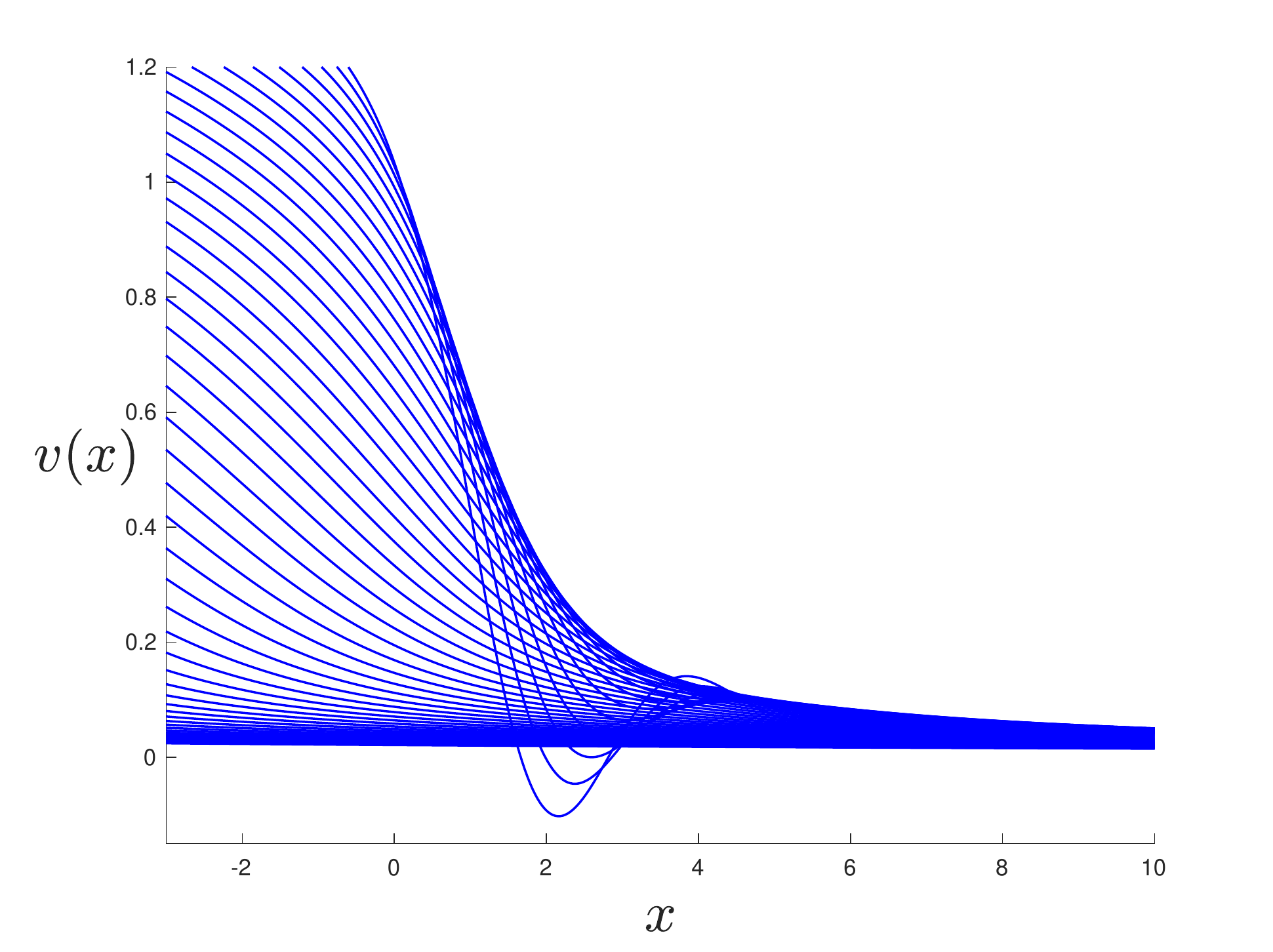}
\caption{\label{fig:Supplement-D} }
\end{figure}

As before, solutions can be readily generated for smaller $\hbar$, and indeed the result is  reduced room to explore the $s<0$ region before the rise begins. In other words  reducing $\hbar$ gets closer to the semi-classical expectations, as should be the case. These are all important lessons for the full JT gravity case.

\section{III. The JT gravity case}
As explained in the Letter, JT gravity can be built as a sum of   the  models for all $k$, and so it potentially inherits some of the behaviour of the $k$ even models. What happens~\cite{Saad:2019lba} is that the semi-classical $V_{\rm eff}(s){=}(4\pi^3\hbar)^{-1}[\sin(2\pi\sqrt{-s}){-}2\pi\sqrt{-s}\cos(2\pi\sqrt{-s})]$ does not go negative indefinitely, but oscillates  back and forth between positive and negative values, with increasing magnitude,  as $s$ goes more negative. All of those regions where $V_{\rm eff}(s)$ is negative will result in a peak in the distribution, and since there are an infinite number of them, the problem is again expected to degenerate, destroying the classical configuration. However, there are be finite $\hbar$ corrections to give the true $V_{\rm eff}(s)$ but it is not clear (and seems unlikely) that such corrections can erase all the infinite number of zeroes. The numerical explorations done here were difficult to take to low enough $s$ to see peaks arising. Perhaps other methods can reveal if $s$ can be taken safely all the way to $-\infty$.

In any case, a sensible and meaningful non-perturbative definition can be given by studying the ensemble of matrices where the lowest energy $s$ is at some finite value, $\sigma$. Choosing $\sigma$ above the place where $V_{\rm eff}(s)$ (fully corrected) first goes negative will ensure that the peak of interest (for example in  figure~\ref{fig:JT-ground-state-distribution}) represents a sensible family of  ground states that connects sensibly to topological perturbation theory of ref.\cite{Saad:2019lba}. The choice $\sigma=0$ was made originally in ref.~\cite{Johnson:2019eik}. Also, for the many supersymmetric cases subsequently studied~(see \cite{Johnson:2022wsr} and references therein), such a choice has extra motivation since then the Hamiltonian is the square of a supercharge.

Finally, figure~\ref{fig:Supplement-E}  shows a family of six $v(x)$ curves for some positive $s$ values, and the corresponding curves for $\psi(x)^2={\cal R}(x)/2$. For the latter, the linear rise discussed in the text is visible, allowing for the Letter's formula for the leading positive $s$ asymptotic to be readily derived. A striking thing about it is its $\hbar$-independence, and the fact that it is built out of the Bessel function relation that directly follows from the leading ``Schwarzian'' spectral density~(\ref{eq:schwarzian-leading}) at large $E$: The universality of the Schwarzian directly translates into the universality of this side of the peak.

\begin{figure}[h]
\centering
\includegraphics[width=0.47\textwidth]{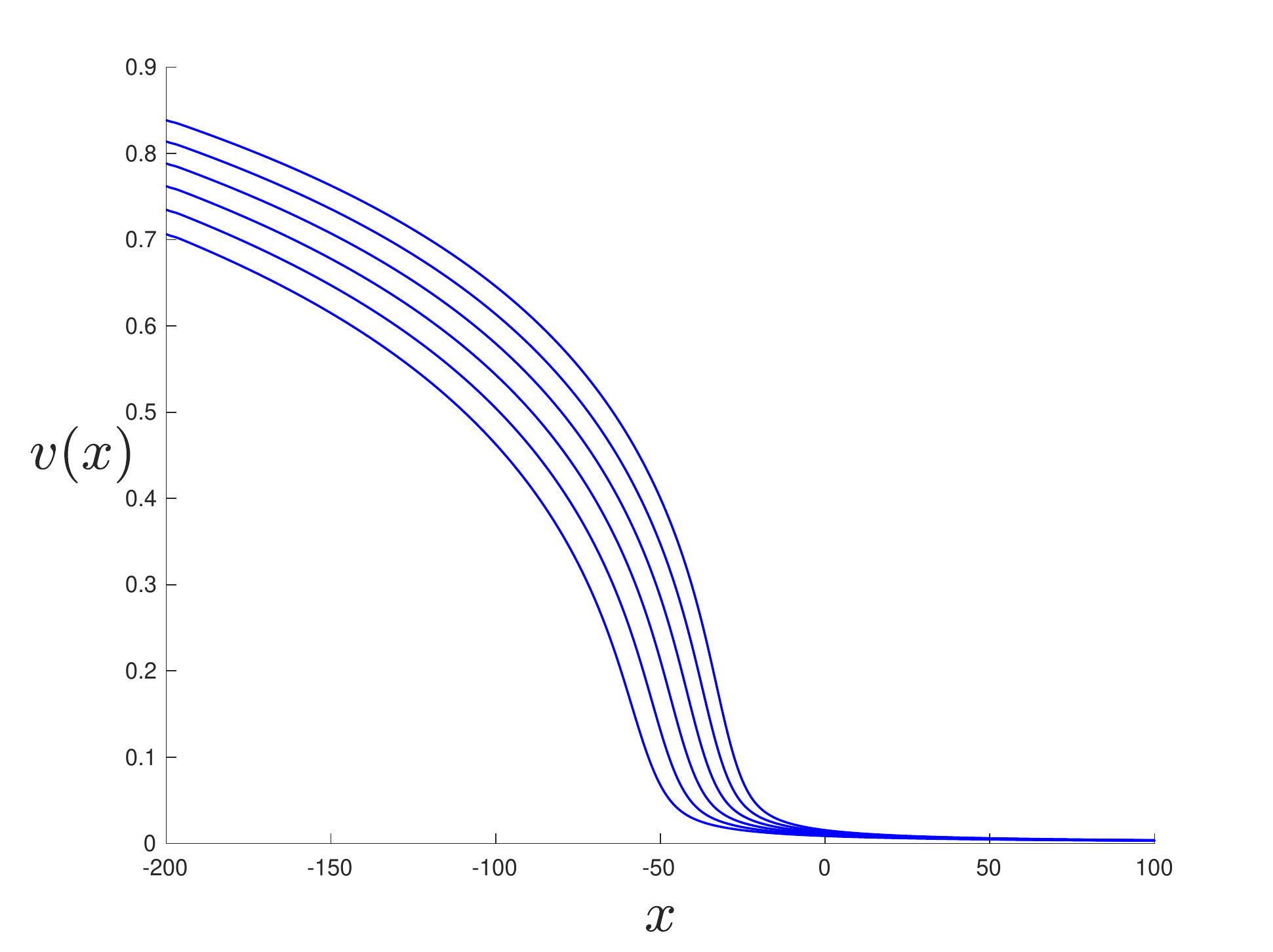}\hskip0.5cm \includegraphics[width=0.45\textwidth]{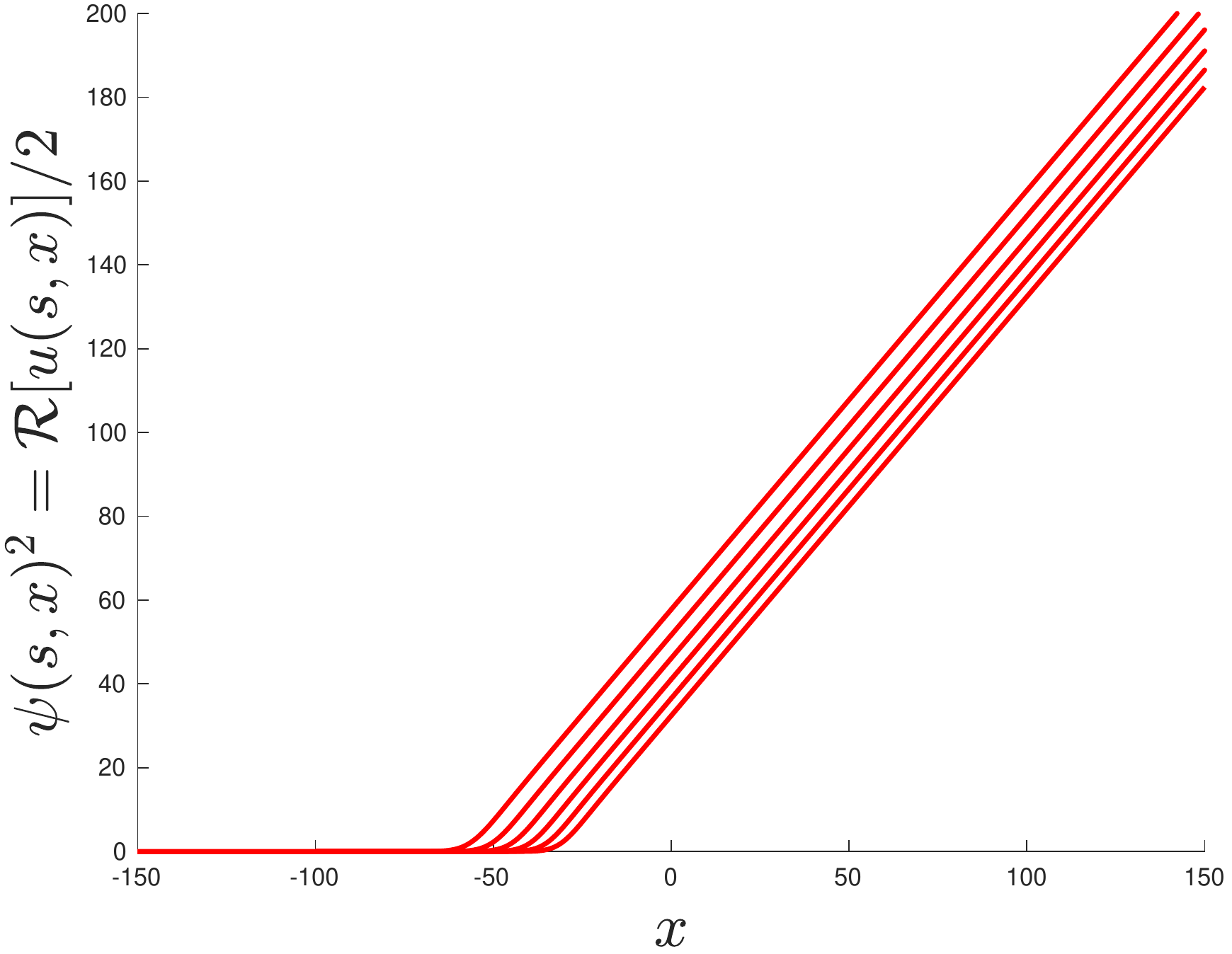}
\caption{\label{fig:Supplement-E} }
\end{figure}

\end{widetext}
\end{document}